# Societal-Scale Decision Making Using Social Networks

Marko Antonio Rodriguez
University of California, Santa Cruz
okram@soe.ucsc.edu

Daniel Joshua Steinbock
University of California, Santa Cruz
daniel@sonic.net

**Abstract**
In societal-scale decision-making, a collective is faced with the problem of deriving a decision that is in accord with the collective's intentions and values. Modern political institutions utilize representational structures for decision-making such that any individual in the society can, in potential, participate in the decision-making behavior of the collective—even if only indirectly through a proxy representative. An agent-based simulation demonstrates that in traditional representation structures, as the size of the total population increases linearly relative to the number of decision-making representatives, there is an exponential increase in the likelihood that decision outcomes will not accurately reflect the preferences of the collective. In the direction of a remedy, this paper describes a novel social network-based method for societal-scale decision-making which greatly improves the accuracy of representative decision outcomes. This work shows promise for the future development of policy-making systems that are supported by the computer and network infrastructure of our society.

Contact:
Marko Antonio Rodriguez
Computer Science Department
University of California, Santa Cruz
Santa Cruz, CA 95060

Tel: 831-459-5625
Email: okram@soe.ucsc.edu

# Societal-Scale Decision-Making Using Social Networks
Marko Antonio Rodriguez and Daniel Joshua Steinbock

In societal-scale decision-making systems, a collective is faced with the problem of deriving a decision that is in accord with the collective's intention. Modern political institutions utilize representational structures for decision-making such that any individual in the society can, in potential, participate in the decision-making behavior of the collective—even if only indirectly through a proxy representative. An agent-based simulation demonstrates that in traditional representation structures, as the size of the total population increases linearly relative to the number of decision-making representatives, there is an exponential increase in the likelihood that decision outcomes will not accurately reflect the preferences of the collective. In the direction of a remedy, this paper describes a novel social network-based method for societal-scale decision-making which greatly improves the accuracy of representative decision outcomes. This work shows promise for the future development of policy-making systems that are supported by the computer and network infrastructure of our society.

## Social Compression and the Overload Problem

Collective decision-making is central to the functioning of a society and supported by a system of rules, procedures, power structures and participants. An *overload problem* occurs when a collective does not have the information-processing infrastructure to support the active participation of all its constituent members in all decision-making processes [Fischer 1999; Rodriguez 2004]. To overcome this issue, societies have come to approximate full participation by using a set of decision-making representatives. This approximation is analogous to a computer scientist's concept of "lossy" data compression where some loss of information is tolerated in order to reduce the resources required for storage or communication. Accordingly, we can call the use of representative decision-makers *social compression* and measure the amount of information loss as the ratio between those being represented to those representing. A lossless 1-to-1 representational structure is the case when all individuals are representatives of themselves, a direct democracy. At the other extreme, when the ratio of representation reaches an all-to-1 model, one individual is the autocratic representative of all members in the group. This lowest-resolution representation structure is a gross lossy model of the group since the ability to represent the perspective of every individual becomes increasingly difficult as the size and diversity of the group increases [Rodriguez 2004].

Most modern democratic institutions lie in between these two extremes, a regime where the number of active decision-makers is large enough to represent large-scale trends in public opinion but small enough to keep communications overhead manageable. As a rough illustration, the current resident population of the United States is estimated by the U.S. Census Bureau to be approximately 293 million (as of 2004) while the size of U.S. Congress is fixed by law at 535 members. This gives a representation ratio of approximately one policy-maker per 547,000 citizens. Presumably, there is a congressional membership limit because the communications overhead required to conduct traditional parliamentary process has a practical ceiling; yet there is also the simple fact that the architecture of the congressional meeting chamber permits only a limited number of seats. As human population increases, and the ratio of representation grows more severe, such artificial constraints on the policy-making infrastructure of a society become increasingly disabling.

This paper considers and contrasts "traditional" representative decision-making with a newly proposed method based on the dynamic delegation of proxy power across a social network. This new method increases the likelihood that decision outcomes will accurately reflect the opinions of the whole population. First we describe the proposed method and give its mathematical formalization. Next we present the results of an agent-based simulation of both the traditional and proposed methods. This is followed by a discussion of implications for the future development of societal-scale decision-making systems that are supported by the computer and network infrastructure of our society.

## Model Description

In this section we describe a simple computer model of collective decision-making in order to compare two alternative forms of representation: 1) a traditional form in which representatives' opinions are weighted equally when making a decision; 2) a novel form in which the underlying social network of the collective is used to adjust the relative weight of representatives' opinions. As we will show, this latter method is more likely to accurately reflect the opinions of the whole collective.

The simulation defines an individual as a node within a social network. Each individual node is assigned an "opinion" value from a uniform distribution between 0.0 and 1.0. Figuratively, one could imagine a node with a 0.0 opinion as an extreme conservative and an individual with a 1.0 opinion as an extreme liberal. Values in between these bounds represent the diverse opinions of the general population. In the equations to follow, the set $N$ denotes the entire population while the subset $A$ ($A \subseteq N$) denotes the active participants (representatives). Equation (1) defines a decision outcome as the average of representatives' opinion values. Equation 2 gives the expected decision outcome when the whole collective participates ($A = N$); this is the standard by which we measure how accurately the decision made by a subset (1) reflects the opinions of the whole (2).

$$GroupDecision = \frac{1}{|A|} \sum_{p \in A} opinion(p) \qquad (1)$$

$$ExpectedDecision = \frac{1}{|N|} \sum_{p \in N} opinion(p) \qquad (2)$$

Intuitively, the closer the number of actively participating individuals (|A|) is to the size of the total population (|N|), the more accurately the group is able to model the perspective of all its constituent members. The decision error of the group is determined by the absolute value of the difference between the calculated group decision (1) and the expected decision (2). Equation 1 is a complete description of the first form of representative decision-making we are considering where the opinions of participants are treated equally to produce an outcome. The second form (3) is identical except that participants' opinions are unequally weighted in order to more accurately reflect the opinions of non-participants. The *weight(p)* parameter in (3) is the number of non-participants being represented by participant p (including p itself). This weight may be fractional (non-integer) but the sum of all weights always equals the total number of participants, $N$. Thus equation (3) is simply a weighted average of participants' opinions.

$$WeightedGroupDecision = \frac{1}{|N|} \sum_{p \in A} weight(p) * opinion(p) \qquad (3)$$

For our simulation, the method we choose to assign these weights is to presuppose the existence of a social network which represents trust relationships among members of the collective. The following equation defines the amount of trust individual $p$ has in individual $q$:

$$Trust(p,q) = 1 - |opinion(p) - opinion(q)| \qquad (4)$$

According to (4) similarity of opinions implies a symmetric trust relationship. This is an oversimplification of trust formation in the real world which is often asymmetric and based on factors other than shared opinions such as relevant expertise. However this equation serves the purposes of our simulation by providing a plausible basis for social network relationships. Trust values are assigned as edge-weights on the directed edges of the social network. Notice that the range of (4) is [0,1] so edge-weights in our model actually denote the *percentage* of each node's trust assigned to each adjacent node.

The only thing that remains to be defined is the method for calculating *weight(p)* values in (3) based on the network topology and edge-weights. As an illustration, we'll describe the weight calculation method in the context of a realistic social network with asymmetric trust assignments. Figure 1 below depicts a simple example showing trust connections among a four node collective with two active participants (shown by stars). The weight of each active node's opinion when calculating decision outcomes in (3) follows directly from the trust given them by non-active nodes. The great utility of this social network method is that we take advantage of trust transitivity. Decision power travels along paths of trust, automatically delegating to the active participants in a natural way. In figure 1, human A trusts B completely and B divides trust unequally between C and D. We can imagine that each node is initially given one unit of trust which will be divided among its peers according to edge-weights. The process iterates with each node redistributing trust received on the previous iteration—except for active nodes which only collect trust and do not redistribute. This continues until all trust has been aggregated to the set of active nodes. Note how this implies a kind of conservation of energy. Formal algorithms for this aggregation process are presented by [Steinbock 2004]. Another important topic which is not covered in this paper is that, since we trust different people for different reasons, individuals would need a different set of peers for each subject domain of decision-making (see the discussion of organizational domain modeling in [Rodriguez 2004]).

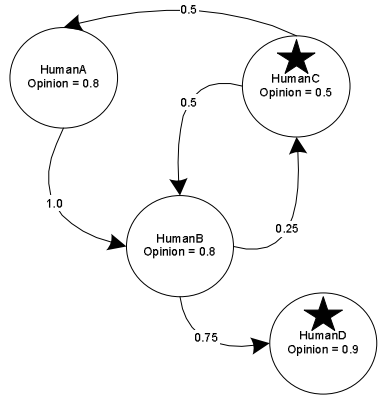

**Fig. 1: Social trust network of representatives (stars)**

For the example illustrated in figure above, we apply (2) to get an expected decision of 0.75 [(0.9+0.8+0.8+0.5)/4]. C's weight is 1.5 [1.0 + .25 + .25], while D's weight is 2.5 [1.0 + .75 + .75]. According to the first decision method given by (1), the outcome would be 0.7 [(.9 + .5)/2]; compared against the expected decision this gives an error of 0.05. If we instead use the social network method (3), the outcome would be 0.75 [((0.9 * 2.5) + (0.5 * 1.5))/4] giving us zero error for this example.

Now that a preliminary understanding of the social network method has been presented, we summarize the results of our simulation runs on networks of one hundred with constant connectivity (# of outgoing edges). Our intent was to measure the accuracy of group decision outcomes relative to expected outcomes and compare the traditional method to our social network method. The results in figure 2 clearly show the advantage of our method for varying ratios of representation. While we also studied the effect of network connectivity, this parameter merely scaled the measured advantage by a constant factor. For clarity we only show data for network connectivity of three ($K \approx 3$), which exhibited the highest performance compared to the traditional method.

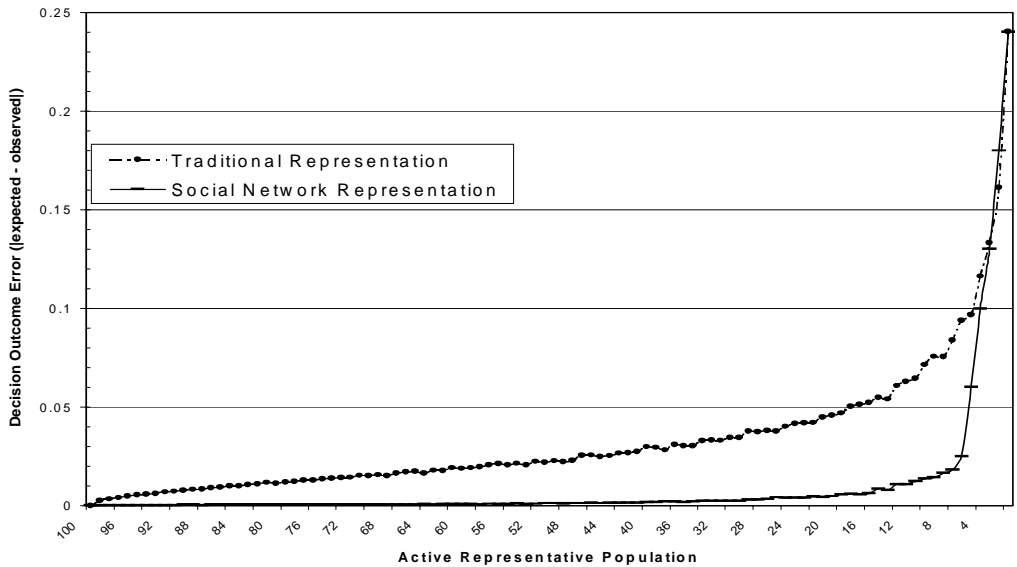

**Figure 2: Accuracy of traditional representation vs. social network-based method**

As expected, supplementing the traditional method with a social trust network for weighting the opinions of representatives resulted in a large decrease in decision error (as measured against the expected outcome). This result was especially dramatic when the active population was small relative to total population. The intuitive reason we see this result is that the use of a social network dampens the effect of a particular choice of representatives. Whereas the traditional method makes choosing representatives critical to the decision outcome—especially when the set is small—the social network smoothes out fluctuations so that a relatively stable model of the collective opinion is maintained, regardless of which particular subset of the population is making decisions for the whole.

## Discussion and Concluding Remarks

Due to the overload problem, collective decisions are often made by a subset of the population; with respect to a given decision, this amounts to partitioning the collective into two sets: participants and non-participants. Participants are at minimum representatives of their own opinions but in practice represent the entire collective—insomuch that decisions determine collective actions. We have considered a traditional method of representational decision-making where decision outcomes derive solely from the opinions of the participating individuals. According to our simulation, this method results in an exponential increase in decision error as the number of representatives decreases linearly relative to the size of the total population (figure 2). However, by bringing to bear the knowledge implicit in a social network of trust relationships, the simulation indicates that this increase in error can be significantly dampened for nearly any number of representatives and any particular choice of representatives. This research has important implications for collectives whose availability of human resources fluctuates rapidly while the structure of the underlying social network is relatively stable. It offers a way to maintain a relatively stable approximation of collective opinion using nearly *any* subset of the population as representatives. This is analogous to a hologram, where any broken-off part of the whole image is in fact a lower-resolution version of the whole.

The idea of dynamic representation has an important role to play in the future development of societal-scale decision-making systems as public policy-making becomes more embedded within the medium of the world's modern network and computer infrastructure [Heylighen 2002; Turoff 2002]. The increasing complexity and interconnectedness of global society makes decentralization both necessary and attainable; formally, this complexity transition corresponds to a shift from hierarchical control structures to participatory networks [Bar-Yam 1997]. It is our position that dynamic representation is a critical part of this shift as it plays out in the context of public policy-making. In order to manage the complexity of global society, it will be necessary to replace the traditionally static, hierarchical forms of representation with new network-based models which adapt to the rapidly changing dynamics and contexts of decentralized society. It is our hope that future designers of large-scale human decision-making systems will find our social networks-based method of use in meeting this emerging need.